\documentclass[10pt]{iopart}
\usepackage{epsfig,cite}
\newcommand{\beq}{\begin{equation}}
\newcommand{\eeq}{\end{equation}}
\newcommand{\ds}{\displaystyle}
\newcommand{\beqar}{\begin{eqnarray}}
\newcommand{\eeqar}{\end{eqnarray}}

\begin{document}

\title[Strangeness at SPS and RHIC within two-source model]{
Strangeness production in heavy ion collisions at SPS and RHIC 
within two-source statistical model}
\author{
Z-D~Lu\dag\ddag, Amand~Faessler\ddag, C~Fuchs\ddag,
E~E~Zabrodin\ddag\S
}
\address{\dag\
China Institute of Atomic Energy, Beijing 102413, China}
\address{\ddag\
Institut f\"ur Theoretische Physik, Universit\"at
    T\"ubingen, T\"ubingen, Germany}
\address{\S\
Institute for Nuclear Physics, Moscow State University, Moscow, 
    Russia}

\begin{abstract}
The experimental data on hadron yields and ratios in central Pb+Pb
and Au+Au collisions at SPS and RHIC energies, respectively, are 
analysed within a two-source statistical model of an ideal hadron gas. 
These two sources
represent the expanding system of colliding heavy ions, where the hot
central fireball is embedded in a larger but cooler fireball. The
volume of the central source increases with rising bombarding energy.
Results of the two-source model fit to RHIC experimental data at
midrapidity coincide with the results of the one-source thermal model
fit, indicating the formation of an extended fireball, which is
three times larger than the corresponding core at SPS.
\end{abstract}

\vspace{-0.5cm}
\section{Introduction. Two-source model}
\vspace{-0.1cm}
Searching for the quark-gluon plasma (QGP) is one of the main goals
in the study of relativistic heavy ion collisions. The principal
question is whether the strongly interacting matter reaches the stage 
of chemical and thermal equilibrium.
Although this idea was put forward by Fermi 50 years ago \cite{Fer50},
up to now there is no unambiguous test to probe the degree of
equilibration in the system. One of the possible approaches is to
study the equilibration process within microscopic models
\cite{GeKa93,Brplb98,Brprc99,SHSX99}.
The more traditional way is to fit yields and transverse spectra of 
particles obtained in experiments to the statistical model (SM) of a 
fully equilibrated hadron gas (see 
\cite{Raf91,Bec96,YeGo99,BHS99,ClRe99,LeRa99,Lu98,Zha00,BrMrhic} and 
references therein). In the model the macroscopic characteristics of 
the system, such as the particle number density, the energy density, 
etc., are derived via a set of distribution functions (we assume 
$c = \hbar = k_B = 1$)
\beq
\ds
f(p,m_i) = \left\{ \exp{\left[ \left(\sqrt{p^2 + m_i^2} -\mu_B B_i -
\mu_S S_i\right) /T \right] }\pm 1 \right\} ^{-1}\ ,
\label{eq1}
\eeq
where $p$ and $m_i$ are the momentum and the mass of the hadron
species $i$, $T$ is the temperature, $\mu_B$ and $\mu_S$ are the
baryon chemical potential and strangeness chemical potential, $B_i$
and $S_i$ are the baryon charge and strange charge of hadron $i$. The
sign ``+" in Eq.~(\ref{eq1}) is for fermions and sign``$-$" 
for bosons. 

As the experimental data became more precise, it has been understood
that the ideal SM does not provide an adequate description of all
hadron multiplicities. Particularly, the yields of pions are usually 
underestimated while the abundances of strange particles are 
overpredicted. Therefore, some modifications to the SM have been 
proposed, such as excluded volume effects \cite{YeGo99,BHS99};
strangeness suppression, $\gamma_S < 1$ \cite{Raf91,Bec96};
chemical non-equilibrium of light quarks \cite{LeRa99}.
The assumption of a single expanding source is the basic 
{\it ad hoc\/} hypothesis of these models. 
However, if the baryon
density and/or the strangeness density are not the same everywhere in
the reaction volume, the scenario with several independent sources
cannot be reduced to the single source scenario. Our investigations of
the two-source scenario are inspired by the experimental observation 
\cite{na52} and microscopic model predictions \cite{Brprc99} of low net 
baryon densities in the midrapidity range of relativistic heavy-ion 
collisions at SPS energies. Thus, local equilibrium 
may occur separately in different zones of the fireball.

The proposed two-source statistical model (TSM) of a hadron gas 
\cite{tsm} divides the whole reaction zone into two
regions: the outer region (source 1 or S1) and the inner region
(source 2 or S2). The two sources are allowed to possess different
temperatures, net baryon and strangeness densities, etc.
The characteristics of each of the fireballs can be described via the 
four independent parameters, such as volume $V$, fireball 
temperature $T$, net baryon density $\rho_B$, and net strangeness 
density $\rho_S$. Compared to the SM, the number of free parameters in 
the two-source model increases to seven: Although the net strangeness 
in each of the sources can be nonzero, they are linked via the the 
condition of total strangeness conservation
$ \ds N_{S1} + N_{S2}  = 0\, ,$ i.e., either $\rho_{S1}$ or
$\rho_{S2}$ may be considered as a free parameter.

\vspace{-0.1cm}
\section{Hadron production at SPS and RHIC}
\vspace{-0.1cm}
The baryon yield and ratios of hadrons at midrapidity in central Pb+Pb
collisions at 158 AGeV are listed in Table~\ref{tab1} together with
the results of the TSM and the SM fit.
All hadrons with masses less than 2 GeV/$c^2$ are included in the 
fitting procedure. No additional constraints such as strangeness 
suppression or excluded volume are assumed except the feeding-back 
effect from resonance decay. Compared to the ideal SM, the TSM
improves the agreement with the experimental data.
The thermodynamic quantities obtained from the two fits to
experimental data are shown in Table~\ref{tab1} also.
One can see that the two-source object can be interpreted as
a hot, relatively small core surrounded by a cooler and larger halo.
The major part of baryons is contained in the outer source, while the 
inner source contains almost all antibaryons. 

\begin{table}
\caption{
Baryon yield and hadron ratios at midrapidity for central lead-lead
collisions at SPS energies and predictions of the single-source and
two-source statistical models of an ideal hadron gas.
\label{tab1}}
\begin{tabular}{@{}lllllllll}
\br
   & Data & SM & TSM & Coll. &  & S1 & S2 & SM \\
\mr
$N_B\ (net)$               & 372$\pm 10$  & 371.9 & 372.1& NA49 &
 $T [MeV]$ & 117 & 155 & 158 \\
$K^+/K^- $                 & 1.85$\pm$ 0.1 & 1.97  & 1.87 & NA44 &
 $V [fm^3]$ & 7250 & 1705 & 4203 \\
$\overline{p}/p$           & 0.07$\pm$ 0.01 & 0.069 & 0.058 & NA44  &
 $N_B$ & 348 & 55  & 408 \\
$\overline{\Xi}/\Xi$       & 0.249$\pm$ 0.019 & 0.231 & 0.247 & WA97 &
 $N_{\overline{B}}$ & 0 & 30 & 36 \\
$\overline{\Omega}/\Omega$ & 0.383$\pm$ 0.081 & 0.427 & 0.405 & WA97 &
 $N_S$ & 34 & 42 & 117 \\
$\overline{\Lambda}/\Lambda$ & 0.128$\pm$ 0.012& 0.130 & 0.137 & WA97 &
 $N_{\overline{S}}$ & 21 & 55 & 117 \\
$\eta/\pi^0$               & 0.081$\pm$0.013& 0.133 & 0.108  & WA98  &
 $\mu_B [MeV]$ & 460 & 45 & 213 \\
$K_S^0/\pi^-$              & 0.125$\pm$0.019& 0.121 & 0.120  & NA49  &
 &  &  & \\
$K_S^0/h^-$                & 0.123$\pm$0.02 & 0.102 & 0.102  & WA97   &
 &  &  & \\
$\Lambda/h^-$              & 0.077$\pm$0.011& 0.069 & 0.064  & WA97 &
  &  &  &  \\
$\Omega/\Xi$               & 0.219$\pm$0.045& 0.131 & 0.104  & WA97 &
  &  &  &  \\
$\Xi^-/\Lambda$            & 0.110$\pm$0.01 & 0.156 & 0.115  & WA97 &
  &  &  &  \\
$\chi^2/DOF$                 &                 & 46/9 & 16/5 & &
  &  &  & \\
\br
\end{tabular}
\end{table}

The strangeness density is negative in S2 and positive in S1, i.e.,
the inner source contains more $s$-quarks than $\bar{s}$-quarks. This 
finding is supported by microscopic model calculations \cite{Brprc99}. 
The possible explanation of the phenomenon is as follows: 
strange and anti-strange particles must be produced in pairs. 
Because of the small
interaction cross section with hadrons, $K^+$ and $K^0$ are leaving
the central reaction zone easier than strange particles which carry
$s$ quarks, $\Lambda$ and $\overline{K}$, thus leading to a
negative strangeness density in the midrapidity range.
 
The energy density in S2 is about three times larger than that in
S1. Such a low energy density in the outer source corresponds to
the energy density at thermal freeze-out rather than at chemical
freeze-out \cite{Brprc99}. In other words, the solution for two
sources cannot be reduced to the one-source picture even in the
case where exclusively midrapidity data have been used.
Similar fit has been performed to solely $4 \pi$-data and to a 
mixture of $4 \pi$-data with midrapidity data. Results of all three 
fits favour the idea of the formation of a compact hot baryon-dilute
central zone with the
following averaged characteristics: temperature $T = 157 \pm 2$ MeV,
volume $V = 0.6 \pm 0.1\ V_0$, where $V_0$ is the volume of a lead 
nucleus, and baryon chemical potential $\mu_B = 31 \pm 14$ MeV. 
The temperature of the halo is much lower, $T_{S_1} = 117 \pm 3$ MeV.

Experimental data on hadron yields and ratios in the midrapidity
range in central Au+Au collisions at $\sqrt{s} = 130$ AGeV
are listed in Table~\ref{tab2} together with the
predictions of the SM and TSM. Surprisingly, now the 
results of the SM fit and the TSM fit are almost identical.
It seems that the volume of the central fireball significantly
increases by the transition from the SPS energies to the RHIC ones,
and that hadrons detected in the midrapidity region are originated
from a single thermalized source. Its volume is more than 5000 fm$^3$,
i.e., three times larger than the core volume at SPS,
and the temperature reaches 186 MeV. 
However, if the multiplicity of negatively charged
hadrons, $h^-$, is excluded from the data set, the temperature of
the modelled system drops to 176 MeV, that is extremely close to the
value $T = 175$ MeV obtained in \cite{BrMrhic}. This
important question should be clarified in future studies. We checked
that the incorporation of the excluded volume effects by
assigning the hard-core radius $r = 0.4$ fm to all particles leads
to an enlargement of the volume but does not affect the
temperature of the fireball.

Another useful characteristics at the chemical freeze-out are
the energy per particle and the entropy per baryon. In \cite{ClRe99}
the criterion $E/N \approx 1$ GeV was introduced.
The predictions of the SM and
the TSM for the midrapidity range are as follows: $E/N = 1.1$ GeV,
and $S/A \equiv s/\rho_B =121$. The last value is about 20\%
below the value $s/\rho_B =150$ predicted by the UrQMD calculations,
but the latter are for Au+Au collisions at $\sqrt{s} = 200$ AGeV
\cite{LVrhic}. It would be interesting to perform these microscopic
model calculations also at $\sqrt{s} = 130$ AGeV.

\begin{table}
\caption{
The same as Table~\ref{tab1} but for Au+Au collisions at 
$\sqrt{s} = 130$ AGeV.
\label{tab2}}
\begin{tabular}{@{}lllllllll}
\br
   & Data & SM & TSM & Coll. &  & SM & S1 & S2 \\
\mr
$N_B\ (net)$              & 343$\pm 11$    & 340.6 & 340.6 & PHOBOS & 
 $T [MeV] $         & 185.7 & 185.9 & 185.5 \\
$h^-$                      &2050$\pm 250$   &2238   & 2239  & PHOBOS &
 $V [fm^3]$         & 5157  & 2608  & 2560  \\
$\overline{p}/p$           & 0.60$\pm$ 0.07 & 0.57  & 0.57 & PHOBOS &
 $N_B$              & 893   & 456   & 437   \\
$\overline{p}/\pi^-$       & 0.08$\pm$ 0.01 & 0.087 & 0.087 & STAR   &
 $N_{\overline{B}}$ & 553   & 281   & 272   \\
$K^-/K^+ $                 & 0.87$\pm$ 0.08 & 0.75  & 0.75  & STAR   &
 $N_S$              & 307   & 155   & 152   \\
$K^-/\pi^-$                & 0.149$\pm$0.02 & 0.153 & 0.153 & STAR   &
 $N_{\overline{S}}$ & 307   & 155   & 152   \\
$K^{\ast 0}/h^-$           & 0.060$\pm$0.017& 0.036 & 0.036 & STAR   &
 $\mu_B [MeV]$      & 53    & 53    & 52    \\
$\overline{K^{\ast 0}}/h^-$ & 0.058$\pm$0.017& 0.030 & 0.030 & STAR  &
  &  &  & \\
$\overline{\Lambda}/\Lambda$ & 0.77$\pm$ 0.07& 0.69 & 0.69  & STAR   &
  &  &  & \\
$\overline{\Xi}/\Xi$       & 0.82$\pm$ 0.08 & 0.80  & 0.80  & STAR   &
  &  &  & \\
$\chi^2/DOF$               &                & 9.6/3  & 9.6/7 & \\
\br
\end{tabular}
\end{table}

\vspace{-0.1cm}
\section{Conclusions}
\vspace{-0.1cm}
In summary, 
the TSM fit to the experimental data taken at midrapidity at RHIC
coincides with the standard single-source fit. This
result supports the idea of a formation of an extended hot fireball
in the central zone of heavy-ion collisions at RHIC energies.
The temperature of the central fireball varies from 176 MeV to
185 MeV, depending on the incorporation of the multiplicity of
negatively charged hadrons in the fit, whereas the
excluded volume effects seem not to affect the fireball temperature.

At SPS it is found that the properties of the system at chemical
freeze-out can be well understood in terms of two sources, a central
core and a surrounding halo, both being in local chemical and thermal
equilibrium. Temperatures as well as baryon charge and strange
charge of the two sources are different.

Strangeness seems to be in equilibrium in both sources. 
This observation is in line with the fact
that $\gamma_S \approx 1$ if one fits the particle ratios from the
midrapidity range to the SM
\cite{BHS99}, but $\gamma_S < 1$ if one intends to fit $4\pi$-data
\cite{Bec96,YeGo99}. A possible explanation for this puzzle is a
non-homogeneous distribution of the strange charge within the
reaction volume. 
It would be interesting to study the forthcoming 
$4 \pi$-data on Au+Au collisions at RHIC energies in order to check 
(i) the increase of the volume of the central fireball w.r.t.
the halo; (ii) equilibration of strangeness in both sources; 
(iii) a possible change of the halo temperature.  

{\bf Acknowledgements.}
We are grateful to L. Bravina for fruitful discussions.
The work was supported by the DFG, BMBF under the contract 
No. 06T\"U986, and the NSF of China under the contracts
No. 19975075 and No. 19775068.

\end{document}